\documentclass[proceedings]{JHEP3}
\usepackage{eurosym}
\usepackage{amsfonts}
\usepackage{amsmath}
\usepackage{epsfig}

\newbox\mybox

\newcommand\fverb{\setbox\mybox=\hbox\bgroup\verb}
\newcommand\fverbdo{\egroup\medskip\noindent\fbox{\unhbox\mybox}\ }
\newcommand\fverbit{\egroup\item[\fbox{\unhbox\mybox}]}
\conference{Spectrally equivalent t-dependent double wells/unstable anharmonic oscillators}
\abstract{We construct a time-dependent double well potential as an exact spectral equivalent to the explicitly time-dependent negative quartic oscillator with a time-dependent mass term. 
Defining the unstable anharmonic oscillator Hamiltonian on a contour in the lower-half complex plane, the resulting time-dependent non-Hermitian Hamiltonian is first mapped by an exact solution of the
time-dependent Dyson equation to a time-dependent Hermitian Hamiltonian defined on the real axis. When unitary transformed, scaled and Fourier transformed we obtain 
a time-dependent double well potential bounded from below. All transformations are carried out non-perturbatively so that all Hamiltonians in this process are
spectrally exactly equivalent in the sense that they have identical instantaneous energy eigenvalue spectra.}

\title{Spectrally equivalent time-dependent double wells and unstable
anharmonic oscillators}
\author{Andreas Fring and Rebecca Tenney \\
Department of Mathematics, City University London,\\
Northampton Square, London EC1V 0HB, UK\\
E-mail: a.fring@city.ac.uk, rebecca.tenney@city.ac.uk}

\input{tcilatex}
\begin{document}

\section{Introduction}

Anharmonic oscillators have a wide range of applications in quantum
mechanics as they describe for instance delocalization and decoherence of
quantum states, e.g. \cite{milburn1986}. They also occur naturally in
relativistic models, e.g. \cite{gabrielse1985}. From a mathematical point of
view their nonlinear nature make them ideal testing grounds for various
approximation methods, such as perturbative approaches \cite{seznec1979}.
Based on a perturbative expansion of the energy eigenvalues it was shown in 
\cite{andrianov1982} that the quartic anharmonic oscillator with mass term
is spectrally equivalent to a double well potential with linear symmetry
breaking. The first hint about the fact that even the unstable quartic
anharmonic oscillator posses a well defined bounded real spectrum, despite
being unbounded from below on the real axis, was proved in \cite%
{graffi1983,caliceti1988}, where it was proven that its energy eigenvalues
series is Borel summable. The spectral equivalence between an unstable
anharmonic oscillator and a complex double well potential was then proven
directly by Buslaev and Grecchi in \cite{buslaev1993}.

Subsequently the unstable quartic anharmonic oscillator without mass term
was treated in \cite{Bender:1998ke} as part of the general series of $%
\mathcal{PT}$-symmetric potentials $V(x)=x^{2}(ix)^{\varepsilon }$, i.e. $%
\varepsilon =2$, where it was shown numerically that the Hamiltonians in
this series have real and positive spectra for $\varepsilon \geq 2$ .
Applying the techniques developed in this area of non-Hermitian $\mathcal{PT}
$-symmetric quantum mechanics \cite{Alirev,PTbook} Jones and Mateo \cite{JM}
showed that the two Hamiltonians%
\begin{equation}
H=p^{2}-gx^{4},\qquad \text{and\qquad }h=\frac{p^{4}}{64g}-\frac{1}{2}%
p+16gx^{2},  \label{JonesH}
\end{equation}%
are spectrally equivalent. This was established by first defining $H$ on a
suitable contour in the complex plane, $x\rightarrow -2i\sqrt{1+ix}$, within
the Stoke wedges where the corresponding wavefunctions decay asymptotically.
Subsequently the resulting complex Hamiltonian was similarity transformed to
a Hermitian Hamiltonian $h$ that is well defined on the real axis.

Here our central aim is to extend the analysis by making the Hamiltonian
explicitly time-dependent $H\rightarrow H(t)$ through the inclusion of an
explicit time-dependence into the coefficients. The similarity
transformation acquires then the form 
\begin{equation}
h(t)=\eta (t)H(t)\eta ^{-1}(t)+i\partial _{t}\eta (t)\eta ^{-1}(t),
\label{Dyson}
\end{equation}%
often referred to as the time-dependent Dyson equation \cite%
{CA,time1,time6,BilaAd,time7,fringmoussa,AndTom1,AndTom3,most2018en}, in
which $H\neq H^{\dagger }$ is a non-Hermitian explicitly time-dependent
Hamiltonian, $h=h^{\dagger }$ a Hermitian explicitly time-dependent
Hamiltonian and\ $\eta (t)$ the time-dependent Dyson map. The latter can be
used to define a time-dependent metric $\rho (t)$ via the relation $\rho
(t)=\eta ^{\dagger }(t)\eta (t)$. Spectral equivalence is then understood on
the level of the instantaneous energy eigenvalues for the operators $h(t)$
and the corresponding operator for the non-Hermitian system 
\begin{equation}
\tilde{H}(t)=\eta ^{-1}(t)h(t)\eta (t)=H(t)+i\eta ^{-1}(t)\partial _{t}\eta
(t).  \label{energy}
\end{equation}%
Note while $\tilde{H}$ is observable it is not a Hamiltonian governing the
time-evolution and satisfying the time-dependent Schr\"{o}dinger equation.
On the other hand the Hamiltonian $H(t)$ is not observable. Besides the
aforementioned interest in the unstable anharmonic oscillator itself, there
are not many known exact solutions \cite%
{BilaAd,fringmoussa,fringmoussa2,AndTom1,AndTom2,AndTom3,khantoul2017invariant,maamache2017pseudo,cen2019time,AndTom4,AndTom5,fring2019eternal,fring2020time,koussa2020pseudo}
to the time-dependent Dyson equation (\ref{Dyson}), so that any new exact
solution provides valuable insights.

\section{The time-dependent unstable harmonic oscillator}

The Hamiltonian we investigate here is similar to the one in equation (\ref%
{JonesH}), but with time-dependent coefficient functions and an additional
mass term 
\begin{equation}
H(z,t)=p^{2}+\frac{m(t)}{4}z^{2}-\frac{g(t)}{16}z^{4},~~~~~m\in \mathbb{R}%
\text{,}g\in \mathbb{R}^{+}\text{.}  \label{tanharm}
\end{equation}%
Defining $H(z,t)$ now on the same contour in the lower-half complex plane $%
z=-2i\sqrt{1+ix}$ as suggested by Jones and Mateo \cite{JM}, it is mapped
into the non-Hermitian Hamiltonian 
\begin{equation}
H(x,t)=p^{2}-\frac{1}{2}p+\frac{i}{2}\{x,p^{2}\}-m(t)(1+ix)+g(t)(x-i)^{2},
\label{H2}
\end{equation}%
with $\{\cdot ,\cdot \}$ denoting the anti-commutator. Next we attempt to
solve the time-dependent Dyson equation (\ref{Dyson}) to find a Hermitian
counterpart $h$. Making the following general Ansatz for the Dyson map%
\begin{equation}
\eta (t)=e^{\alpha (t)x}e^{\beta (t)p^{3}+i\gamma (t)p^{2}+i\delta
(t)p},~~~~~\alpha ,\beta ,\gamma ,\delta \in \mathbb{R}\text{,}  \label{eta}
\end{equation}%
we use the Baker-Campbell-Hausdorff formula to compute the adjoint action of 
$\eta (t)$ on all terms appearing in $H(x,t)$%
\begin{eqnarray}
\eta x\eta ^{-1} &=&x+\delta +6\alpha \beta p+2\gamma p+3i\alpha ^{2}\beta
+2i\alpha \gamma -3i\beta p^{2},  \label{ad1} \\
\eta p\eta ^{-1} &=&p+i\alpha , \\
\eta x^{2}\eta ^{-1} &=&x^{2}-9\beta ^{2}p^{4}-12i\beta \left( 3\alpha \beta
+\gamma \right) p^{3}+\left( 54\alpha ^{2}\beta ^{2}+36\alpha \beta \gamma
+4\gamma ^{2}-6i\beta \delta \right) p^{2} \\
&&+4(3\alpha \beta +\gamma )\left[ \delta +i\alpha (3\alpha \beta +2\gamma )%
\right] p+2\left( \delta +3i\alpha ^{2}\beta +2i\alpha \gamma \right) x 
\notag \\
&&+(6\alpha \beta +2\gamma )\left\{ x,p\right\} -3i\beta \left\{ x\text{,}%
p^{2}\right\} -(3\alpha ^{2}\beta +2\alpha \gamma -i\delta )^{2},  \notag \\
\eta p^{2}\eta ^{-1} &=&p^{2}-\alpha ^{2}+2i\alpha p, \\
\eta \{x,p^{2}\}\eta ^{-1} &=&\{x,p^{2}\}-6i\beta p^{4}+(24\alpha \beta
+4\gamma )p^{3}+\left( 36i\alpha ^{2}\beta +12i\alpha \gamma +2\delta
\right) p^{2}-2\alpha ^{2}x~\ \ ~~~~  \label{ad6} \\
&&+4\left( i\alpha \delta -6\alpha ^{3}\beta -3\alpha ^{2}\gamma \right)
p-2i\alpha ^{2}\left( 3\alpha ^{2}\beta +2\alpha \gamma -i\delta \right)
+4i\alpha \text{$\{$}x,p\text{$\}.$}  \notag
\end{eqnarray}%
The gauge like terms in (\ref{Dyson}) and (\ref{energy}) are calculated to%
\begin{eqnarray}
i\dot{\eta}\eta ^{-1} &=&ix\dot{\alpha}+\allowbreak \allowbreak i\dot{\beta}%
p^{3}-\left( 3\dot{\beta}\alpha +\dot{\gamma}\right) p^{2}-\left( 3i\dot{%
\beta}\alpha ^{2}+2i\dot{\gamma}\alpha +\dot{\delta}\right) p+\allowbreak 
\dot{\beta}\alpha ^{3}+\dot{\gamma}\alpha ^{2}-i\dot{\delta}\alpha ,~~\ 
\label{gterm} \\
i\eta ^{-1}\dot{\eta} &=&ix\dot{\alpha}+i\dot{\beta}p^{3}-(3\dot{\alpha}%
\beta +\dot{\gamma})p^{2}-(2i\gamma \dot{\alpha}+\dot{\delta})p-i\delta \dot{%
\alpha},  \label{gt2}
\end{eqnarray}%
where as commonly used we abbreviate partial derivatives with respect to $t$
by an overdot. Using the expressions in (\ref{ad1})-(\ref{gterm}) for the
evaluation of (\ref{Dyson}) and demanding the right hand side to be
Hermitian yields the following constraints for the coefficient functions in
the Dyson map 
\begin{equation}
\alpha =\frac{\dot{g}}{6g},~~~\beta =\frac{1}{6g},~~~\gamma =\frac{%
12g^{3}+6mg^{2}+\dot{g}^{2}-g\ddot{g}}{4\dot{g}g^{2}},~~~\delta =c_{1}\frac{g%
}{\dot{g}}-\frac{g\ln g}{2\dot{g}},  \label{const}
\end{equation}%
with $c_{1}\in \mathbb{R}$ being an integration constant. Moreover, the
time-dependent coefficient functions in the Hamiltonian (\ref{tanharm}) must
be related by the third order differential equation%
\begin{equation}
9g^{2}\left( \dddot{g}-6g\dot{m}\right) +36g\dot{g}\left( gm-\ddot{g}\right)
+28\dot{g}^{3}=0.  \label{third}
\end{equation}%
Integrating once and introducing a new parameterization function $\sigma (t)$%
, we solve this equation by%
\begin{equation}
g=\frac{1}{4\sigma ^{3}},\quad ~~\text{and\quad ~~}m=\frac{4c_{2}+\dot{\sigma%
}^{2}-2\sigma \ddot{\sigma}}{4\sigma ^{2}},  \label{mg}
\end{equation}%
with $c_{2}\in \mathbb{R}$ denoting the integration constant corresponding
to the only integration we have carried out. The time-dependent Hermitian
Hamiltonian in equation (\ref{Dyson})~then results to%
\begin{equation}
h(x,t)=\sigma
^{3}p^{4}+f_{pp}(t)p^{2}+f_{x}(t)x+f_{p}(t)p+f_{xp}(t)\{x,p%
\}+f_{xx}(t)x^{2}+C(t).
\end{equation}%
with%
\begin{eqnarray*}
f_{pp} &=&\frac{\sigma \left\{ \sigma \left[ 2\left( \sigma \left( \dot{%
\sigma}^{2}-4c_{2}\right) -2\right) \ddot{\sigma}+16c_{2}^{2}+\dot{\sigma}%
^{4}\right] +16c_{2}\right\} +4}{4\sigma \dot{\sigma}^{2}},~~f_{xp}=\frac{%
\left( \sigma \left( \dot{\sigma}^{2}-4c_{2}\right) -2\right) }{4\sigma ^{2}%
\dot{\sigma}},~~ \\
f_{p} &=&\frac{2c_{1}\left[ \sigma \left( 4c_{2}+\dot{\sigma}^{2}-2\sigma 
\ddot{\sigma}\right) +2\right] +\ln \left( 4\sigma ^{3}\right) }{12\sigma 
\dot{\sigma}^{2}},~~f_{x}=-\frac{2c_{1}+\ln \left( 4\sigma ^{3}\right) }{%
12\sigma ^{2}\dot{\sigma}},~\ f_{xx}=\frac{1}{4\sigma ^{3}},~ \\
C &=&\frac{\left( 2c_{1}+\ln \left( 4\sigma ^{3}\right) \right) {}^{2}+36%
\dot{\sigma}^{2}\left( 4c_{2}^{2}+\ddot{\sigma}\right) }{144\sigma \dot{%
\sigma}^{2}}+\frac{1}{8}\left( \dot{\sigma}^{2}-4c_{2}\right) \ddot{\sigma}-%
\frac{\dot{\sigma}^{2}}{4\sigma ^{2}}
\end{eqnarray*}%
We may choose to set $c_{1}=c_{2}=0$ and reintroduce the original
time-dependent coefficient functions $g(t)$, $m(t)$ so that the Hamiltonian
simplifies to 
\begin{eqnarray}
h(x,t) &=&\frac{p^{4}}{4g}+\left( \frac{18g^{2}(2g+m)}{\dot{g}^{2}}+\frac{%
\dot{g}^{2}}{72g^{3}}-\frac{2g+m}{4g}\right) p^{2}-\frac{3\left(
g^{2}m+g^{3}\right) \ln g}{\dot{g}^{2}}p+\frac{g^{2}\ln (g)}{\dot{g}}x~~ 
\notag \\
&&+\left( \frac{\dot{g}}{12g}-\frac{6g^{2}}{\dot{g}}\right) \text{$\{x,p\}$}%
+gx^{2}+\frac{1296g^{8}\ln ^{2}g+\dot{g}^{6}-36\dot{g}^{4}g^{2}(2g+m)}{%
5184g^{5}\dot{g}^{2}}-\frac{m}{2}.
\end{eqnarray}%
Notice that $\sigma (t)$ can be any function, but the coefficient functions $%
g(t)$ and $m(t)$ must be related by (\ref{third}) that is (\ref{mg}).

The massless case for $m(t)=0$ is more restrictive and leads to $\sigma (t)$
being a second order polynomial $\sigma (t)=\kappa _{0}+\kappa _{1}t+\kappa
_{2}t^{2}$ with real constants $\kappa _{i}$. \ This case is consistently
recovered from (\ref{mg}) with the choice $c_{2}=\kappa _{1}\kappa
_{3}-\kappa _{2}^{2}/4$. The solution found for the time-independent case in 
\cite{JM}, would be obtained from (\ref{eta}) in the limits $\alpha
\rightarrow 0$, $\beta \rightarrow 1/6g$, $\gamma \rightarrow 0$, $\delta
\rightarrow i$ and $m\rightarrow 0$. While this limit obviously exists for $%
\alpha $ and $\beta $, the constraints for $\gamma $ and $\delta $ are
different from those reported in (\ref{const}). In fact, setting $\delta
(t)\rightarrow i\delta (t)$ enforces $g$ to be time-independent and there is
no time-dependent solution corresponding to that choice. The energy operator 
$\tilde{H}$ defined in (\ref{energy}) is obtained directly by adding $H(x,t)$
in (\ref{H2}) and the gauge-like term in (\ref{gt2}).

Let us now eliminate the terms in $h(x,t)$ proportionate to $x$ and $\{x,p\}$
by means of a unitary transformation 
\begin{equation}
U=e^{-i\frac{f_{xp}}{2f_{xx}}p^{2}-i\frac{f_{x}}{2f_{xx}}p},
\end{equation}%
which leads to the unitary transformed Hamiltonian%
\begin{equation}
\hat{h}(x,t)=\sigma ^{3}p^{4}+\left( f_{pp}-\frac{f_{xp}^{2}}{f_{xx}}\right)
p^{2}+\left( f_{p}-\frac{f_{x}f_{xp}}{f_{xx}}\right) p+f_{xx}x^{2}+C-\frac{%
f_{x}^{2}}{4f_{xx}}.
\end{equation}%
Similarly as in the time-independent case \cite{JM}, we may scale this
Hamiltonian, albeit now with a time-dependent function, $x\rightarrow
(f_{xx})^{-1/2}x$. Subsequently we Fourier transform $\hat{h}(x,t)$ so that
it is viewed in momentum space. In this way we obtain a spectrally
equivalent Hamiltonian with a time-dependent potential%
\begin{eqnarray}
\tilde{h}(y,t) &=&p_{y}^{2}+\sigma ^{3}f_{xx}^{2}y^{4}+\left(
f_{xx}f_{pp}-f_{xp}^{2}\right) y^{2}+\left( \sqrt{f_{xx}}f_{p}-\frac{%
f_{x}f_{xp}}{\sqrt{f_{xx}}}\right) y+C-\frac{f_{x}^{2}}{4f_{xx}},~~ \\
&=&\frac{g}{4}y^{2}\left( y^{2}+\frac{\dot{g}^{2}}{36g^{3}}+\frac{72g^{2}m}{%
\dot{g}^{2}}-\frac{m}{g}+2\right) +\frac{\left( 36g^{2}m+\dot{g}^{2}\right) 
\sqrt{g}\ln g}{12\dot{g}^{2}}y  \label{bounded} \\
&&+\frac{\dot{g}^{4}}{5184g^{5}}-\frac{\dot{g}^{2}m}{144g^{3}}-\frac{\dot{g}%
^{2}}{72g^{2}}-\frac{m}{2},  \notag
\end{eqnarray}%
where for simplicity we have set $c_{1}=c_{2}=0$ in (\ref{bounded}). The
potential in $\tilde{h}(y,t)$ is a double well that is bounded from below.
We illustrate this for a specific choice of $\sigma (t)$, that is $g(t)$ and 
$m(t)$, in figure 1.

\FIGURE{ \epsfig{file=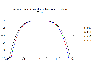, width=7.8cm} \epsfig{file=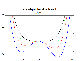,width=6.6cm}
\caption{Spectrally
equivalent time-dependent anharmonic oscillator potential $V(z,t)$ in (\ref{tanharm}) and time-dependent double well potential $\tilde{V}(y,t)$ in (\ref{bounded}) for 
$\sigma (t)=\cosh t$, $g(t)=1/4\cosh ^{3}t$, $m(t)=(\tanh ^{2}t-2)/4$ at
different values of time.}
       \label{Fig1}}

\section{Conclusions}

We have proven the remarkable fact that the time-dependent unstable
anharmonic oscillator is spectrally equivalent to a time-dependent double
well potential that is bounded from below. The transformations we carried
out are summarized as follows: 
\begin{equation*}
H(z,t)\overset{z\rightarrow x}{\rightarrow }H(x,t)\overset{\text{Dyson}}{%
\rightarrow }h(x,t)\overset{\text{unitary transform}}{\rightarrow }\hat{h}%
(x,t)\overset{\text{Fourier}}{\rightarrow }\tilde{h}(y,t).
\end{equation*}%
We have first transformed the time-dependent anharmonic oscillator $H(z,t)$
from a complex contour in a Stokes wedge to the real axis $H(x,t)$. The
resulting non-Hermitian Hamiltonian $H(x,t)$ was then mapped by mean of a
time-dependent Dyson map $\eta (t)$ to a time-dependent Hermitian
Hamiltonian $h(x,t)$. It turned out that the Dyson map can not be obtained
by simply introducing time-dependence into the known solution for the
time-independent case \cite{JM}, but it required to complexify one of the
constants and the inclusion of two additional factors. In order to obtain a
potential Hamiltonian we have unitary transformed $h(x,t)$ into a spectrally
equivalent Hamiltonian $\hat{h}(x,t)$, which when Fourier transformed leads
to $\tilde{h}(y,t)$ that involved a time-dependent double well potential.

A detailed analysis of the spectra and eigenfunctions using approximation
methods for time-dependent potential \cite{BeckyAnd1} is left for future
investigations. Moreover, it is well known that Dyson maps are not unique,
in the time-dependent as well as time-independent case, and it might
therefore be interesting to explore whether additional spectrally equivalent
Hamiltonians to $H(z,t)$ can be found in the same fashion for new type of
maps.\medskip 

\noindent \textbf{Acknowledgments:} RT is supported by a City, University of
London Research Fellowship.

\newif\ifabfull\abfulltrue

%%\bibliographystyle{phreport}
%%\bibliography{acompat,Ref}

\end{document}